\documentclass[12pt]{article}
\usepackage[dvips]{graphicx}
\usepackage{amsmath,amssymb,amsthm, color}
\def\frk{\mathfrak}               

\def\Phi{{\frk N}}
\def\opn#1#2{\def#1{\operatorname{#2}}} 
\opn\chara{char} 
\opn\length{\ell} 
\opn\pd{pd} 
\opn\rk{rk}
\opn\projdim{proj\,dim} 
\opn\injdim{inj\,dim} 
\opn\rank{rank}
\opn\depth{depth} 
\opn\grade{grade} 
\opn\height{height}
\opn\embdim{emb\,dim} 
\opn\codim{codim}

\opn\Tr{Tr} 
\opn\bigrank{big\,rank}
\opn\superheight{superheight}
\opn\lcm{lcm}
\opn\trdeg{tr\,deg}
\opn\reg{reg} 
\opn\lreg{lreg} 
\opn\ini{in} 
\opn\lpd{lpd}
\opn\size{size}
\opn\mult{mult}
\opn\dist{dist}
\opn\cone{cone}
\opn\lex{lex}
\opn\rev{rev}
\opn\div{div} \opn\Div{Div} \opn\cl{cl} \opn\Cl{Cl}
\opn\Spec{Spec} \opn\Supp{Supp} \opn\supp{supp} \opn\Sing{Sing}
\opn\Ass{Ass} \opn\Min{Min}
\opn\Ann{Ann} \opn\Rad{Rad} \opn\Soc{Soc}
\opn\Syz{Syz} \opn\Im{Im} \opn\Ker{Ker} \opn\Coker{Coker}
\opn\Am{Am} \opn\Hom{Hom} \opn\Tor{Tor} \opn\Ext{Ext}
\opn\End{End} \opn\Aut{Aut} \opn\id{id} \opn\ini{in}

\opn\nat{nat}
\opn\pff{pf}
\opn\Pf{Pf} \opn\GL{GL} \opn\SL{SL} \opn\mod{mod} \opn\ord{ord}
\opn\Gin{Gin}
\opn\Hilb{Hilb}\opn\adeg{adeg}\opn\std{std}\opn\ip{infpt}
\opn\Pol{Pol}
\opn\sat{sat}
\opn\Var{Var}
\opn\Gen{Gen}

\opn\aff{aff} \opn\con{conv} \opn\relint{relint} \opn\st{st}
\opn\lk{lk} \opn\cn{cn} \opn\core{core} \opn\vol{vol}
\opn\link{link} \opn\star{star}
\opn\gr{gr}

\def\Ac{{\mathcal A}}

\def\Jc{{\mathcal J}}
\def\Gc{{\mathcal G}}

\def\Rc{{\mathcal R}}

\def\pot#1#2{#1[\kern-0.28ex[#2]\kern-0.28ex]}

\opn\dirlim{\underrightarrow{\lim}}
\opn\inivlim{\underleftarrow{\lim}}
\let\to=\rightarrow

\def\Implies{\ifmmode\Longrightarrow \else
        \unskip${}\Longrightarrow{}$\ignorespaces\fi}
\def\implies{\ifmmode\Rightarrow \else
        \unskip${}\Rightarrow{}$\ignorespaces\fi}
\def\iff{\ifmmode\Longleftrightarrow \else
        \unskip${}\Longleftrightarrow{}$\ignorespaces\fi}

\let\:=\colon
\newtheorem{Theorem}{Theorem}[section]
\newtheorem{Lemma}[Theorem]{Lemma}

\theoremstyle{definition}

\newtheorem{Remark}[Theorem]{Remark}

\newtheorem{Example}[Theorem]{Example}
\newtheorem{Definition}[Theorem]{Definition}

\let\epsilon\varepsilon
\let\phi=\varphi
\let\kappa=\varkappa
\def\qed{\ifhmode\textqed\fi
      \ifmmode\ifinner\quad\qedsymbol\else\dispqed\fi\fi}
\def\textqed{\unskip\nobreak\penalty50
       \hskip2em\hbox{}\nobreak\hfil\qedsymbol
       \parfillskip=0pt \finalhyphendemerits=0}
\def\dispqed{\rlap{\qquad\qedsymbol}}

\def\Bt{{\bf t}}
\def\Bx{{\bf x}}

\def\By{{\bf y}}

\def\supp{{\rm supp}}

\def\Bp{{\bf p}}

\def\Bt{{\bf t}}

\def\Bz{{\bf z}}
\def\Bm{{\bf m}}

\def\Bone{{\bf 1}}
\def\Bzero{{\bf 0}}

\title{
Markov bases for 
two-way change-point models of\\
ladder determinantal tables
}
\author{Satoshi Aoki and Takayuki Hibi}

\begin{document}
\maketitle

\begin{abstract}
To evaluate the goodness-of-fit of a statistical model to given data, 
calculating a conditional $p$ value by a Markov chain Monte Carlo method
 is one of the effective approaches. For this purpose, a Markov basis
 plays an important role because it 
 guarantees the connectivity of the chain, which is needed for
 unbiasedness of the
 estimation, and therefore is investigated in various settings such as
 incomplete tables or subtable sum constraints. In this paper, we
 consider the two-way change-point model for the ladder determinantal
 table, which is an extension of these two previous works, i.e., 
works on incomplete tables by Aoki and Takemura (2005, {\it
 J. Stat. Comput. Simulat.}) and subtable
 some constraints by Hara, Takemura and Yoshida (2010, {\it J. Pure
 Appl. Algebra}). Our main
 result is based on the theory of Gr\"obner basis for the distributive
 lattice. We give a numerical example for actual data.
\end{abstract}

\section{Introduction}
In the analysis of contingency tables, computing conditional $p$ values
by a Markov chain Monte Carlo method is one of the common approaches to
evaluate a fitting of a statistical model to given data. 
In this method, a key notion is a {\it Markov basis} that 
guarantees the connectivity of the chain for
unbiasedness of the estimation. 
In Diaconis and Sturmfels (\cite{Diaconis-Sturmfels-1998}), 
a notion of a Markov basis is presented with 
algebraic algorithms to compute it.
This first work is based on a discovery of 
the relation between a Markov basis and a set of binomial generators of
a toric ideal of a polynomial ring, which 
is the first
connection between commutative algebra and statistics.
After this first paper, Markov bases are studied
intensively by many researchers both in the fields of commutative
algebra and statistics, which yields an attractive new field called {\it
computational
algebraic statistics}. See \cite{Pistone-Riccomagno-Wynn-2001} for the
first textbook of this field, and 
\cite{Aoki-Hara-Takemura-2012} for various
theoretical results and examples on Markov bases. 

The first result on the Markov bases in 
the setting of two-way contingency tables
 is a Markov basis for the independence model. 
For two-way
contingency tables with fixed row sums and column sums, which is the
minimal sufficient statistics under the independence model, the set of
square-free moves of degree $2$ forms a Markov basis. This result is
generalized to the decomposable models of higher dimensional contingency
tables by \cite{Dobra-2003}. The reader can find 
various results on the structure of Markov bases of decomposable models
in Chapter 8 of \cite{Aoki-Hara-Takemura-2012}. 

On the other hand, it is known that 
the structure of a Markov basis becomes complicated 
under various additional constraints to the two-way setting. 
One of such cases 
is the {\it incomplete two-way contingency table}, i.e., a contingency table
with {\it structural zeros}, considered in \cite{Aoki-Takemura-2005}.
Another case is the {\it subtable sum problem} considered in 
\cite{Hara-Takemura-Yoshida-2010} and \cite{TWOWAY}. 
In these works, it is shown that 
moves of higher degrees are needed for Markov bases.
The problem we consider in this paper is two-way contingency tables with 
both structural zeros and subtable sum constraints. 

We consider the two-way contingency tables with specific
types of structural zeros called {\it ladder determinantal tables}, with
specific types of subtable sums called {\it two-way change-point model}.
The two-way change-point model is considered in \cite{Hirotsu-1997} for
exponential families, including the Poisson distribution for complete
two-way contingency tables. We also consider the Poisson distribution
and two-way change-point model for incomplete cases in this paper. 
The purpose of this paper is to show that a Markov basis for
this setting is constructed as the set of square-free degree $2$ moves.

This paper is organized as follows. In Section 2, we illustrate the
Markov chain Monte Carlo methods for the subtable sum problem of
incomplete
two-way contingency tables and the two-way change-point models of ladder
determinant tables. In Section 3, we give the structure of the minimal
Markov bases for our problems, which is the main result of this
paper. The arguments and the proof of our main theorem are based on the
theory of Gr\"obner bases for distributive lattices, which is summarized
in Section 3.
A numerical example for actual data is given in Section 4.

\section{Preliminaries}
\subsection{Markov chain Monte Carlo methods for subtable sum problem of
  incomplete contingency tables}
\label{subsec:MCMC-incomplete-table}
First we illustrate the Markov chain Monte Carlo methods for the subtable
sum problem of incomplete two-way contingency tables. Though we only 
consider the two-way change-point model in this paper, we describe the
methods in the setting of general subtable sum problems considered in
\cite{Hara-Takemura-Yoshida-2010}. Note that a specification of the
subtable reduces to the two-way change-point model.

Let $\mathbb{N} = \{0,1,2,\ldots\}$ be the set of nonnegative
integers. To consider $I\times J$ contingency
tables with structural zeros, let $S \subset \{(i,j)\, :\, 1 \leq i \leq
I, 1 \leq j \leq J\}$ be the set of cells that are not structural
zeros. Let $q = |S|$ be the number of the cells. 
Let $\Bx = \{x_{ij}\} \in \mathbb{N}^q$ be an incomplete contingency
table with the set
of cells $S$,
where $x_{ij} \in \mathbb{N}$ is an entry of the cell $(i,j) \in S$. 
Similarly to the ordinary (i.e., complete) two-way contingency tables, 
denote the row sums and column sums of $\Bx$ by
\[
 x_{i+} = \sum_{\{j\, :\, (i,j) \in S\}}x_{ij},\ \ i = 1,\ldots,I,
\]
\[
 x_{+j} = \sum_{\{i\, :\, (i,j) \in S\}}x_{ij},\ \ j = 1,\ldots,J.
\]
We assume that there is at least one $(i,j) \in S$ in
each row and each column. 
Let $B$ be a subset of $S$. 
We also define the subtable sum
$x_B$ by
\[
 x_B = \sum_{(i,j) \in B}x_{ij}.
\]
Denote the set of the row sums, column sums and the subtable sum $x_B$
by an $(I+J+1)$-dimensional column vector 
\begin{equation}
 \Bt = (x_{1+},\ldots,x_{I+},x_{+1},\ldots,x_{+J},x_B)' \in
 \mathbb{N}^{I+J+1},
\label{eqn:minimal-sufficient-t}
\end{equation}
where ${}'$ is the transpose. We also treat $\Bx$ as a $q$-dimensional
column vector as $\Bx = (x_{11},x_{12},\ldots,x_{IJ})'$, by
lexicographic ordering of the cells in $S$. 
Then the relation between $\Bx$ and $\Bt$ is written by
\begin{equation}
 A\Bx = \Bt,
\label{eqn:Ax=t}
\end{equation}
where $A$ is an $(I+J+1) \times p$ matrix consisting of $0$'s and
$1$'s. We call $A$ a {\it configuration matrix}. 
Though we specify $S$ and $B$ in Section \ref{subsec:tables-and-models},
we show an example here. 
\begin{Example}
Consider a $4\times 4$ incomplete contingency table with $6$ structural
 zeros as follows.
\[
 \begin{array}{|c|c|c|c|}\hline
x_{11} & x_{12} & x_{13} & [0]\\ \hline
[0] & x_{22} & x_{23} & x_{24}\\ \hline
[0] & [0] & x_{33} & x_{34}\\ \hline
[0] & [0] & x_{43} & x_{44}\\ \hline
\end{array}
\]
In this paper, we denote a structural zero as $[0]$ to distinguish it
 from a sample zero described as $0$. 
Then the set $S$ is
\[
 S = \{(1,1), (1,2), (1,3), (2,2), (2,3), (2,4), (3,3), (3,4), (4,3), (4,4)\}
\]
and $p = 10$. 
Suppose a subset $B \subset S$ is given by
\[
 B = \{(1,1), (1,2), (1,3), (2,2), (2,3)\}.
\]
Then the configuration matrix is the following $9\times 10$ matrix.
\[
A = \left(
 \begin{array}{cccccccccc}
1 & 1 & 1 & 0 & 0 & 0 & 0 & 0 & 0 & 0\\
0 & 0 & 0 & 1 & 1 & 1 & 0 & 0 & 0 & 0\\
0 & 0 & 0 & 0 & 0 & 0 & 1 & 1 & 0 & 0\\
0 & 0 & 0 & 0 & 0 & 0 & 0 & 0 & 1 & 1\\
1 & 0 & 0 & 0 & 0 & 0 & 0 & 0 & 0 & 0\\
0 & 1 & 0 & 1 & 0 & 0 & 0 & 0 & 0 & 0\\
0 & 0 & 1 & 0 & 1 & 0 & 1 & 0 & 1 & 0\\
0 & 0 & 0 & 0 & 0 & 1 & 0 & 1 & 0 & 1\\
1 & 1 & 1 & 1 & 1 & 0 & 0 & 0 & 0 & 0
\end{array}
\right)
\]
As we see in Section \ref{subsec:tables-and-models}, the
 configuration matrix considered in this
 paper satisfies the homogeneity assumption, i.e., the row vector
 $(1,\ldots,1)$ is in the real vector space spanned by the rows of
 $A$. This is a natural assumption for statistical models. 
See Lemma 4.14 of \cite{Sturmfels-1996} for the algebraic
 aspect of the homogeneity. \hspace*{\fill}$\Box$
\end{Example}

To clarify the statistical meaning of the configuration matrix $A$ and
the relation (\ref{eqn:Ax=t}), 
consider the cell
probability $\Bp = \{p_{ij}\} \in \Delta_{q-1}$, where
\[
 \Delta_{q-1} = \left\{
\{p_{ij}\} \in \mathbb{R}^q_{\geq 0}\, :\, \sum_{(i,j) \in S}p_{ij} = 1
\right\}
\]
is called a $(q-1)$-dimensional probability simplex, and $\mathbb{R}_{\geq
0}$ is the set of nonnegative real numbers.  
The probability simplex $\Delta_{q-1}$ is a statistical model called a
saturated model. In statistical data analysis, our interest is in a
statistical
model that is a subset of $\Delta_{q-1}$. 
The two-way change-point model we consider in this paper is written in
general form by 
\begin{equation}
 {\cal M} = \{\Bp = (p_{ij})\in \Delta_{q-1}\, :\, \log p_{ij} =
 \alpha_i + \beta_j + \gamma\Bone_B(i,j)\ \mbox{for some}\
 (\alpha_i), (\beta_j), \gamma\},
\label{eqn:null-model-M}
\end{equation}
where $\Bone_B(i,j)$ is an indicator function given by
\[
 \Bone_B(i,j) = \left\{\begin{array}{ll}
1, & (i,j) \in B\\
0, & (i,j) \in S \setminus B.
\end{array}
\right.
\]
Here the term $\gamma\Bone_B(i,j)$ represents a departure
from the independence
structure of the log-linear model. 
The model ${\cal M}$ becomes a quasi-independence model for the cells
$S$ by $\gamma = 0$.
The quasi-independence model is a fundamental statistical model for the
incomplete contingency tables (see Chapter 5 of
\cite{Bishop-Fienberg-Holland-1975} for detail). 
Sometimes, the term ``quasi-independence'' is also used for the model of
independence except for the diagonal cells. In this paper, we use the
term ``quasi-independence'' for a larger class of models.
Markov bases for the quasi-independence model are considered in 
\cite{Aoki-Takemura-2005}. Also, the model ${\cal M}$ for the case that 
there are no
structural zeros, i.e., $S = \{1,\ldots,I\} \times \{1,\ldots,J\}$,
corresponds to the setting considered in \cite{Hara-Takemura-Yoshida-2010}. 
The two-way change-point model we consider corresponds to the case
\begin{equation}
 B = \{(i,j) \in S\, :\, i \leq i^*, j \leq j^*\}
\label{eqn:B-two-way-change-point}
\end{equation}
for a fixed $(i^*, j^*) \in S$. 

In this paper, we consider the fitting of the model ${\cal M}$ by the 
statistical hypothesis test
\begin{equation}
 \begin{array}{ll}
\mbox{H}_0:\ \Bp \in {\cal M},\\
\mbox{H}_1:\ \Bp \in \Delta_{p-1}.
\end{array}
\label{eqn:test-problem}
\end{equation}
Under the null hypothesis H$_0$, $(\alpha_i), (\beta_j), \gamma$ in
(\ref{eqn:null-model-M}) are 
nuisance parameters. For testing a null hypothesis in the presence of
nuisance parameters, a common approach is to base the inference on the
conditional distribution given a minimal sufficient statistics for the
nuisance parameters. 
This 
approach is also known as
the Rao-Blackwellization of the
test statistics.
Using this conditional distribution, the
conditional $p$ value is defined. See \cite{Agresti-1992} or Chapter 1 of
\cite{Aoki-Hara-Takemura-2012} for
detail. For our case, the minimal sufficient statistics under the 
null model (\ref{eqn:null-model-M}) is $\Bt = A\Bx$ in
(\ref{eqn:minimal-sufficient-t}), that is the statistical meaning of the
configuration matrix $A$. 
Therefore the conditional distribution under $\mbox{H}_0$, called a {\it
null distribution}, is written by
\[
 f(\Bx\ |\ A\Bx = \Bt) = C^{-1}\prod_{(i,j) \in S}\frac{1}{x_{ij}!},
\]
where $C$ is the normalizing constant written by
\[
 C = \sum_{\By \in {\cal F}_{\Bt}}\left(\prod_{(i,j) \in
 S}\frac{1}{y_{ij}!}\right),
\]
where
\[
 {\cal F}_{\Bt} = \left\{\By \in \mathbb{N}^q\, :\, A\By = \Bt
\right\}.
\]
${\cal F}_{\Bt}$, called a $\Bt$-{\it fiber}, is the set of contingency tables
with given values of row sums, column sums and subtable sum. 
For the observed contingency table $\Bx^o$, 
the conditional $p$ value for the test (\ref{eqn:test-problem})
based on a test statistic $T(\Bx)$ is defined
by
\[
 p = \sum_{\Bx \in {\cal F}_{A\Bx^o}}\phi(\Bx)f(\Bx\ |\ A\Bx = A\Bx^o),
\]
where $\phi(\Bx)$ is the test function of $T(\Bx)$ given by
\[
 \phi(\Bx) = \left\{\begin{array}{ll}
1, & T(\Bx) \geq T(\Bx^o),\\
0, & \mbox{otherwise}.
\end{array}
\right.
\]
To evaluate the conditional $p$ value, a Monte Carlo
approach is to generate samples from the null distribution 
$f(\Bx\ |\ A\Bx = A\Bx^o)$ and calculate the null distribution of the
test statistics. In particular, if a connected Markov chain over 
${\cal F}_{A\Bx^o}$ is constructed, the chain can be modified to give a
connected and aperiodic Markov chain with stationary distribution 
$f(\Bx\ |\ A\Bx = A\Bx^o)$ by a Metropolis procedure, and we can use the
transitions $\Bx^{(M+1)}, \Bx^{(M+2)}, \ldots \in {\cal F}_{A\Bx^o}$ of
the chain after 
a large number of steps $M$, called burn-in steps, as samples from the
null distribution. This is a {\it Markov chain Monte Carlo method}. 
See Chapter 2 of \cite{Aoki-Hara-Takemura-2012} or \cite{Hastings-1970}
for detail.

To construct a connected Markov chain over ${\cal F}_{A\Bx^o}$, 
one of the common approaches is to use a {\it Markov basis} introduced in
\cite{Diaconis-Sturmfels-1998}. An integer array $\Bz \in
\mathbb{Z}^p$ satisfying $A\Bz = \Bzero$ is called a {\it move} for the
configuration $A$, where $\mathbb{Z}$ is the set of integers. Let
\[
 {\cal F}_0(A) = \{\Bz \in \mathbb{Z}^p\, :\, A\Bz = \Bzero\}
\] 
denote the set of moves for $A$. 
\begin{Definition}[\cite{Diaconis-Sturmfels-1998}]
A Markov basis for $A$ is a finite set of moves ${\cal B} =
 \{\Bz_1,\ldots,\Bz_L\} \subset {\cal
 F}_0(A)$ such that, for any $\Bt \in \mathbb{N}^{I+J+1}$ and $\Bx, \By
 \in {\cal F}_{\Bt}$ , there exist $N > 0, (\varepsilon_1,
 \Bz_{\ell_1}),\ldots,(\varepsilon_N, \Bz_{\ell_N}) \in {\cal
 B}$ with $\varepsilon_n \in \{-1,1\}$ such that
\[
 \By = \Bx + \sum_{s = 1}^N\varepsilon_s\Bz_{\ell_s}\ \ \mbox{and}\ \
 \Bx + \sum_{s = 1}^n\varepsilon_s\Bz_{\ell_s} \in {\cal F}_A\
 \mbox{for}\ 1\leq n\leq N.
\]
\end{Definition}
We also define the minimality and uniqueness of the Markov basis.\\
\begin{Definition}
A Markov basis ${\cal B}$ is minimal if no proper subset of ${\cal B}$
 is a Markov basis. A minimal Markov basis is unique if all minimal
 Markov bases differ only by sign changes of the elements.  
\end{Definition}
The fundamental results on 
 uniqueness and minimality of 
Markov bases are given in Chapter 5 of \cite{Aoki-Hara-Takemura-2012}.
For the independence model of the complete $I\times J$ contingency
tables, where the minimal sufficient statistics $A\Bx$ is the row sums
and column sums, 
it is known that the set of square-free moves of degree $2$,
\[
 {\cal B} = \{\Bz(i_1,i_2;j_1,j_2),\ \ 1\leq i_1 < i_2 \leq I, 1 \leq j_1
 < j_2\leq J\},
\]
where $\Bz(i_1,i_2;j_1,j_2) = \{z_{ij}\} \in {\cal F}_0(A)$ is given by
\begin{equation}
 z_{ij} = \left\{\begin{array}{rl}
1, & (i,j) = (i_1,j_1), (i_2,j_2),\\
-1, & (i,j) = (i_1,j_2), (i_1,j_2),\\
0, & \mbox{otherwise}
\end{array}
\right.
\label{eqn:basic-move-elements}
\end{equation}
is a unique minimal Markov basis. 
The square-free moves of degree $2$ above, displayed as
\[
 \begin{array}{r|r|r|}
\multicolumn{1}{c}{} & \multicolumn{1}{c}{j_1} &
 \multicolumn{1}{c}{j_2}\\ \cline{2-3}
i_1 & 1 & -1\\ \cline{2-3}  
i_2 & -1 & 1\\ \cline{2-3}  
\end{array}\ ,
\]
is called a {\it basic move}. In the presence of the structural zeros, 
the set of the basis moves is not a Markov basis in general. For
example, 
as shown in \cite{Aoki-Takemura-2005}, 
incomplete tables with structural zeros as the
diagonal elements, moves of degree $3$ displayed as
\[
 \begin{array}{|c|c|c|}\hline
[0] & +1 & -1\\ \hline
-1 & [0] & +1\\ \hline
+1 & -1 & [0]\\ \hline
\end{array}
\]
are needed for Markov bases. Also, as shown in
\cite{Hara-Takemura-Yoshida-2010}, if the subtable sum $x_B$ is fixed
for the patterns such as 
\[
 (i_1,j_1), (i_2,j_2) \in B,\ \ (i_1,j_2), (i_1,j_3), (i_2,j_1),
 (i_2,j_3) \not\in B,
\]
moves such as 
\[
 \begin{array}{c|c|c|c|}
\multicolumn{1}{c}{} & \multicolumn{1}{c}{j_1} & \multicolumn{1}{c}{j_2}
 & \multicolumn{1}{c}{j_3} \\ \cline{2-4}
i_1 & +1 & +1 & -2\\ \cline{2-4}
i_2 & -1 & -1 & +2\\ \cline{2-4}
\end{array}
\]
are needed for Markov bases.  In this paper, we consider a pattern of
structural zeros $S$, called a ladder determinantal table, and 
a subtable pattern (\ref{eqn:B-two-way-change-point}) corresponding to
a two-way change-point model and show that the set of basic
moves forms a unique minimal Markov basis for this setting.

\subsection {Two-way change-point models of ladder determinantal tables}
\label{subsec:tables-and-models}
\begin{figure}
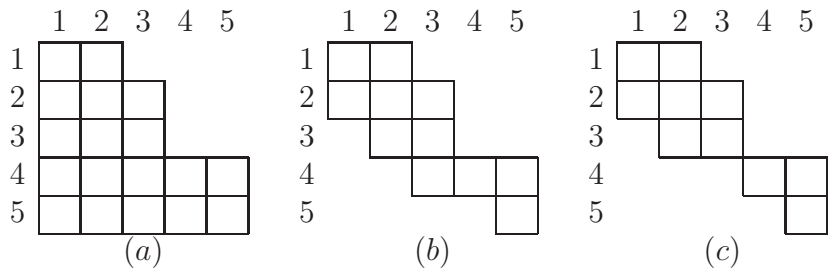

\[
 \begin{array}{c|c|c|c|c|c|}
\multicolumn{1}{c}{} & \multicolumn{1}{c}{1} & \multicolumn{1}{c}{2} &
 \multicolumn{1}{c}{3} & \multicolumn{1}{c}{4} & \multicolumn{1}{c}{5} \\
  \cline{2-3}
1 & & & \multicolumn{3}{c}{}\\ \cline{2-4}
2 & & & & \multicolumn{2}{c}{}\\ \cline{2-4}
3 & & & & \multicolumn{2}{c}{}\\ \cline{2-6}
4 & & & & &  \\ \cline{2-6}
5 & & & & &  \\ \cline{2-6}
\multicolumn{1}{c}{} & 
\multicolumn{5}{c}{(a)} 
\end{array}
\hspace*{5mm}
 \begin{array}{c|c|c|c|c|c|}
\multicolumn{1}{c}{} & \multicolumn{1}{c}{1} & \multicolumn{1}{c}{2} &
 \multicolumn{1}{c}{3} & \multicolumn{1}{c}{4} & \multicolumn{1}{c}{5} \\
  \cline{2-3}
1 & & & \multicolumn{3}{c}{}\\ \cline{2-4}
2 & & & & \multicolumn{2}{c}{}\\ \cline{2-4}
\multicolumn{1}{c}{3} & & & & \multicolumn{2}{c}{}\\ \cline{3-6}
\multicolumn{1}{c}{4} & \multicolumn{1}{c}{} & & & &  \\ \cline{4-6}
\multicolumn{1}{c}{5} & \multicolumn{4}{c|}{} &  \\ \cline{6-6}
\multicolumn{1}{c}{} & 
\multicolumn{5}{c}{(b)} 
\end{array}
\hspace*{5mm}
 \begin{array}{c|c|c|c|c|c|}
\multicolumn{1}{c}{} & \multicolumn{1}{c}{1} & \multicolumn{1}{c}{2} &
 \multicolumn{1}{c}{3} & \multicolumn{1}{c}{4} & \multicolumn{1}{c}{5} \\
  \cline{2-3}
1 & & & \multicolumn{3}{c}{}\\ \cline{2-4}
2 & & & & \multicolumn{2}{c}{}\\ \cline{2-4}
\multicolumn{1}{c}{3} & & & & \multicolumn{2}{c}{}\\ \cline{3-6}
\multicolumn{1}{c}{4} & \multicolumn{2}{c}{} & & &  \\ \cline{5-6}
\multicolumn{1}{c}{5} & \multicolumn{4}{c|}{} &  \\ \cline{6-6}
\multicolumn{1}{c}{} & 
\multicolumn{5}{c}{(c)} 
\end{array}
\]
\caption{Examples of incomplete contingency tables. (a) and (b) are
 ladder determinantal tables, whereas (c) is not.}
\label{fig:example-incomplete-tables}
\end{figure}

Now we specify $S$ considered in this paper.
\begin{Definition}\label{def:ladder-determinantal}
A ladder determinantal table is an incomplete contingency table with the
 set of cells $S \subset
 \{1,\ldots,I\} \times \{1,\ldots,J\}$ satisfying 
\[
 (1,1), (I,J) \in S
\]
and has the form
\begin{equation}
 S = \displaystyle\bigcup_{i = 1}^I\{(i,j),\ \ell_i \leq j \leq u_i\},
\label{eqn:def-ladder-determinantal}
\end{equation}
where $\ell_i \leq \ell_{i+1}, u_i \leq u_{i+1}$ and $u_i \geq \ell_{i+1}$
 hold for $i = 1,\ldots,I-1$.
\end{Definition}
Clearly the condition (\ref{eqn:def-ladder-determinantal}) is also written
by
\[
 S = \displaystyle\bigcup_{j = 1}^J\{(i,j),\ \ell'_j \leq i \leq u'_j\},
\]
where $\ell'_j \leq \ell'_{j+1}, u'_j \leq u'_{j+1}$ and $u'_j \geq \ell'_{j+1}$
 hold for $j = 1,\ldots,J-1$. 
Figure \ref{fig:example-incomplete-tables} 
illustrates examples of incomplete contingency tables.
Figure \ref{fig:example-incomplete-tables}($a$) and
 ($b$) are examples of the ladder determinantal tables, whereas ($c$) is
 not. Figure \ref{fig:example-incomplete-tables}($c$) does not satisfy
 the condition $u_3 \geq \ell_4$ of  
Definition \ref{def:ladder-determinantal} because $u_3 = 3 < 4 = \ell_4$.
\begin{Remark}
The ladder determinantal table above is a special case of 
a {\it block-stairway incomplete table}.
As we see in Chapter 5 of \cite{Bishop-Fienberg-Holland-1975}, an
 incomplete table
 is called a block-stairway table if it is a ladder determinantal table 
after permutation of rows and columns. 
In this paper, we do not consider permutations of rows and columns 
because we consider 
ordered categorical tables. The terminology ``ladder determinantal'' is
 used in algebraic fields. We see the relation between ladder
 determinantal tables and distributive lattices in Section 3. 
\end{Remark}
\begin{Remark}
The condition $u_i \geq \ell_{i+1}$ for $i = 1,\ldots,I-1$ in Definition 
\ref{def:ladder-determinantal} corresponds to the inseparability of
 incomplete tables. See Chapter 5 of
 \cite{Bishop-Fienberg-Holland-1975}. 
We leave this condition because the inseparability is also a natural
 condition in our change-point models. However, it is 
not essential condition in our result, i.e., Theorem \ref{thm:Markov-bases}
 also holds for separable incomplete tables.
\end{Remark}

For the ladder determinantal tables $\Bx$, we consider a 
two-way change-point model, i.e., the model 
(\ref{eqn:null-model-M}) with a subtable $B$ of the form
(\ref{eqn:B-two-way-change-point}).
Though the two-way change-point model is considered in
\cite{Hirotsu-1997} for complete contingency tables, it can be also
considered for incomplete cases. 
We see an example in Section \ref{sec:example}. 

\section{Markov bases of two-way change-point models for ladder
 determinantal tables}
In this section we show the minimal Markov basis for two-way
change-point models for ladder determinantal tables and its uniqueness. 
Note that the set of the
basic moves, i.e, square-free moves of degree $2$, is written by 
\[
 {\cal B}^* = \left\{\Bz(i_1,i_2; j_1,j_2)\ \left|\ 
\begin{array}{cl}
& (i_1,j_1), (i_1,j_2), (i_2,j_1), (i_2,j_2) \in B\\
\mbox{or} & (i_1,j_1), (i_1,j_2) \in B,  (i_2,j_1), (i_2,j_2) \in
 S\setminus B\\
\mbox{or} & (i_1,j_1), (i_2,j_1) \in B,  (i_1,j_2), (i_2,j_2) \in
 S\setminus B\\
\mbox{or} & (i_1,j_1), (i_1,j_2),  (i_2,j_1), (i_2,j_2) \in
 S\setminus B\\
\end{array}
\right.\right\} ,
\]
where $\Bz(i_1,i_2; j_1,j_2) \in {\cal F}_0(A)$ is given by
(\ref{eqn:basic-move-elements}). To show the set ${\cal B}^*$
constitutes a Markov basis, we use the arguments of distributive lattice.

Recall that a {\em partial order} on a set $P$ is a binary relation
$\leq$ on $P$ such that, for all $a, b, c \in P$, one has
\begin{itemize}
\item
$a \leq a$ (reflexivity);
\item
$a \leq b$ and $b \leq a$ $\Rightarrow$ $a = b$ (antisymmetric);
\item
$a \leq b$ and $b \leq c$ $\Rightarrow$ $a \leq c$ (transitivity).
\end{itemize}
A partially ordered set (``poset'' for short) is a set $P$
with a partial order $\leq$.
When $P$ is a finite set, we call $P$ a finite poset.
A {\em lattice} is a poset $L$ for which any two elements $a$ and $b$
belonging to $L$ possess a greatest lower bound (``meet'')
$a \wedge b$ and a least upper bound (``join'')
$a \vee b$. 

\begin{Example}
Let $B_n$ denote the set of subsets 
of $[n] = \{1,2,\ldots,n\}$
and define the partial order $\leq$ on $B_n$ by setting
$X \leq Y$ if $X \subset Y (\subset [n])$.  
Then, in $B_n$, one has $X \cap Y = X \wedge Y$ and
$X \cup Y = X \vee Y$.  Thus $B_n$ is a finite lattice, which is called
the {\em boolean lattice} of rank $n$.
\end{Example}

A lattice is called {\em distributive} if, for all $a, b, c \in L$ one has
\[
a \wedge (b \vee c) = (a \wedge b) \vee (a \wedge c),
\, \, \, \, \, 
a \vee (b \wedge c) = (a \vee b) \wedge (a \vee c).
\]
For example, the boolean lattice of rank $n$ is a distributive lattice.

Let $P$ be a finite poset.  
A {\em poset ideal} of $P$ is a subset $\alpha \subset P$
such that
\[
a \in \alpha, \, \, b \in P, \, \, b \leq a
\, \, \Rightarrow \, \, b \in \alpha.
\]
In particular $P$ itself as well as the empty set $\emptyset$
is a poset ideal of $P$.
Furthermore, if $\alpha$ and $\beta$ are poset ideals of $P$,
then both $\alpha \cap \beta$ and $\alpha \cup \beta$ are 
poset ideals of $P$.

Given a finite poset $P$, we write $L = \Jc(P)$ for the set of
all poset ideals of $P$.  We then define a partial order $\leq$ on $L$
by setting $\alpha \leq \beta$ if $\alpha \subset \beta$,
where $\alpha$ and $\beta$ are poset ideals of $P$.
It follows that $L = \Jc(P)$ is a finite distributive lattice.

A {\em totally ordered subset} of a finite poset $P$ is 
a subset $C$ of $P$ such that, for $a, b \in C$, one has
either $a \leq b$ or $b \leq a$.  
A totally ordered subset of $P$ is also called a {\em chain} of $P$.

Now, a finite distributive lattice $L = \Jc(P)$ 
is called {\em planar} if 
\begin{enumerate}
\item[(i)]
$P$ itself is {\em not} a chain of $P$;
\item[(ii)]
$P$ can be decomposed into the disjoint union 
of two chains of $P$.
\end{enumerate}

\begin{Example}
Let $P = \{a,b,c,d\}$ be a finite poset with
$a < c, b < c, b < d$.  Then $P$ is a disjoint union
of chains $C = \{a, c\}$ and $D = \{b, d\}$.
The finite planar distributive lattice $L = \Jc(P)$ is
Figure \ref{fig:distributive-lattice-L}.
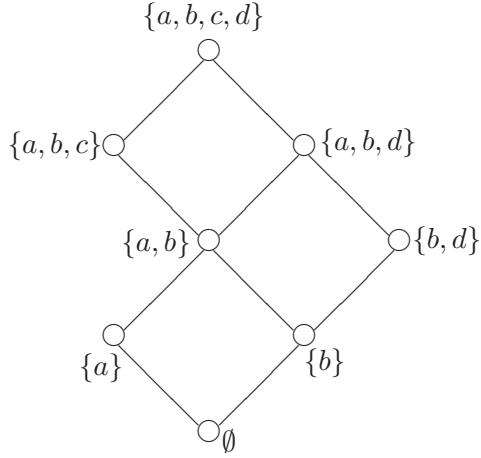
\begin{figure}[htbp]
\begin{center}
\begin{picture}(100,190)(0,0)
\put(50,26){\circle{8}}\put(55,18){{\small $\emptyset$}}
\put(46,28){\line(-1,1){30}}
\put(14,62){\circle{8}} \put(1,47){{\small $\{a\}$}}
\put(54,28){\line(1,1){30}}
\put(86,62){\circle{8}} \put(86,49){{\small $\{b\}$}}
\put(18,64){\line(1,1){30}}
\put(82,64){\line(-1,1){30}}
\put(50,98){\circle{8}} \put(17,94){{\small $\{a,b\}$}}
\put(90,64){\line(1,1){30}}
\put(54,100){\line(1,1){30}}
\put(46,100){\line(-1,1){30}}
\put(14,134){\circle{8}} \put(-26,131){{\small $\{a,b,c\}$}}
\put(18,136){\line(1,1){30}}
\put(50,170){\circle{8}} \put(25,180){{\small $\{a,b,c,d\}$}}
\put(122,98){\circle{8}} \put(127,95){{\small $\{b,d\}$}}
\put(118,100){\line(-1,1){30}}
\put(86,134){\circle{8}} \put(92,132){{\small $\{a,b,d\}$}}
\put(82,136){\line(-1,1){30}}
\end{picture}
\caption{Distributive lattice $L = \Jc(P)$}
\label{fig:distributive-lattice-L}
\end{center}
\end{figure}

\end{Example}

Suppose that $L = \Jc(P)$ is a planar distributive lattice 
for which $P$ is the disjoint union of chains
$C = \{a_{1}, \ldots, a_{n}\}$ and $D = \{b_{1}, \ldots, b_{m}\}$ of $P$
with $a_{1} < \cdots < a_{n}$ and $b_{1} < \cdots < b_{m}$, 
where $n \geq 1$ and $m \geq 1$.
Let 
\[
K[{\bf x}, {\bf y}, s, t] 
= K[x_{1}, \ldots, x_{n}, y_{1}, \ldots, y_{m}, s, t]
\] 
denote the polynomial ring in $n + m + 2$ variables over a field $K$.
We fix a poset ideal $S$ of $L$ with $S \neq \emptyset$ and $S \neq L$.
Given $\alpha \in L$ with
$i_{0} = \max\{ i \, : \, a_{i} \in \alpha \}$ and
$j_{0} = \max\{ j \, : b_{j} \in \alpha \}$,
one can associate the monomial $u_{\alpha} \in K[{\bf x}, {\bf y}, s, t]$ with
\begin{eqnarray*}
u_{\alpha} = \left\{
\begin{array}{ll}
x_{i_{0}}y_{j_{0}}s & \text{if} \, \, \, \alpha \in S, \\
x_{i_{0}}y_{j_{0}}t & \text{if} \, \, \, \alpha \in L \setminus S.
\end{array}
\right.
\end{eqnarray*}
We write $\Rc_{K}[L;S]$ $( \subset K[{\bf x}, {\bf y}, s, t] )$ 
for the toric ring generated by those monomials
$u_{\alpha}$ with $\alpha \in L$.

Let $K[L] = K[z_{\alpha} \, : \, \alpha \in L]$ denote the polynomial ring
in $|L|$ variables over $K$ and fix the reverse lexicographic order $<_{\rm rev}$ 
on $K[L]$ induced by an ordering of the variables of $K[L]$ with the property
that $z_{\alpha} < z_{\beta}$ if $\alpha < \beta$ in $L$. 
We define the surjective ring homomorphism 
$\pi : K[L] \to \Rc_{K}[L;S]$ by setting $\pi(z_{\alpha}) = u_{\alpha}$
with $\alpha \in L$.  
Let $I_{(L;S)}$ $( \subset K[L] )$ denote the kernel of $\pi$,
which will be called the {\em toric ideal} of $\Rc_{K}[L;S]$. 
We refer the reader to, e.g., \cite{dojoEN} for the foundation of 
Gr\"obner bases and toric ideals.

Let $\Ac$ be the set of those $2$-element subsets $\{\alpha, \beta\}$ of $L$,
where $\alpha$ and $\beta$ are incomparable in $L$, satisfying one of
the following:
\begin{itemize}
\item
$\{\alpha, \beta, \alpha \vee \beta\} \subset S$;
\item
$\{\alpha, \beta, \alpha \wedge \beta\} \subset L \setminus S$;
\item
$\alpha \in S$ and $\beta \in L \setminus S$.
\end{itemize}
It then follows that, for each $\{\alpha, \beta\} \in \Ac$, the binomial
\begin{equation}
f_{\alpha, \beta} = z_{\alpha}z_{\beta} - z_{\alpha \wedge
 \beta}z_{\alpha \vee \beta}
\label{eqn:binom-f-a-b}
\end{equation}
belongs to $I_{(L;S)}$.

\begin{Example}
Consider the distributive lattice for 
Table \ref{tbl:hydra-data} we will consider in Section 4. The set
 of the cells of 
Table \ref{tbl:hydra-data} displayed as follows.
\[
 \begin{array}{c|r|r|r|r|r|r|r|}
 \multicolumn{1}{c}{} & \multicolumn{1}{c}{1} & \multicolumn{1}{c}{2} &
 \multicolumn{1}{c}{3} & \multicolumn{1}{c}{4} & \multicolumn{1}{c}{5} &
 \multicolumn{1}{c}{6} & \multicolumn{1}{c}{7}\\ \cline{2-2}
 1 & (1,1) & \multicolumn{6}{|c}{}\\ \cline{2-3}
 2 & (2,1) & (2,2) & \multicolumn{5}{|c}{}\\ \cline{2-4}
 3 & (3,1) & (3,2) & (3,3) & \multicolumn{4}{|c}{}\\ \cline{2-5}
 4 & (4,1) & (4,2) & (4,3) & (4,4) & \multicolumn{3}{|c}{}\\ \cline{2-6}
\multicolumn{1}{c}{5} & \multicolumn{1}{c|}{} & (5,2) & (5,3) & (5,4) & (5,5) &
\multicolumn{2}{|c}{}\\ \cline{3-7}
 \multicolumn{1}{c}{6} & \multicolumn{2}{c|}{} & (6,3) & (6,4) & (6,5) & (6,6) &
\multicolumn{1}{|c}{}\\
  \cline{4-8}
\multicolumn{1}{c}{7} & \multicolumn{3}{c|}{} & (7,4) & (7,5) & (7,6) &
(7,7) \\
  \cline{5-8}
\end{array}
\]
Hereafter we ignore the cells $(1,1)$ and $(7,7)$ because the frequencies
 $x_{11}$ and $x_{77}$ are fixed
 under the model. Then 
the corresponding planar distributive lattice $L$ is displayed in Figure
 \ref{fig:appendix-example}(a). 
In Figure
 \ref{fig:appendix-example}(a),  the set of black vertices $\bullet$
 represents a 
 corresponding poset $P$ where $L = \Jc(P)$, which is also displayed in Figure 
 \ref{fig:appendix-example}(b). 
Note that each vertex $\circ$ or $\bullet$ in Figure 
 \ref{fig:appendix-example}(a) represents a poset ideal of the poset
 consisting of all $\bullet$'s under or equal to it. 
For example, the vertex $\circ$
at 
$(5,4)$ in Figure \ref{fig:appendix-example}(a) represents a poset
ideal 
\[
\{(2,2), (3,1), (3,3), (4,1), (4,4), (5,2)\}, 
\]
of $P$.  
The poset ideal $S \subset L$ 
of Figure \ref{fig:appendix-example}(c) corresponds to 
the two-way change-point model we have considered in Section
 \ref{sec:example}.

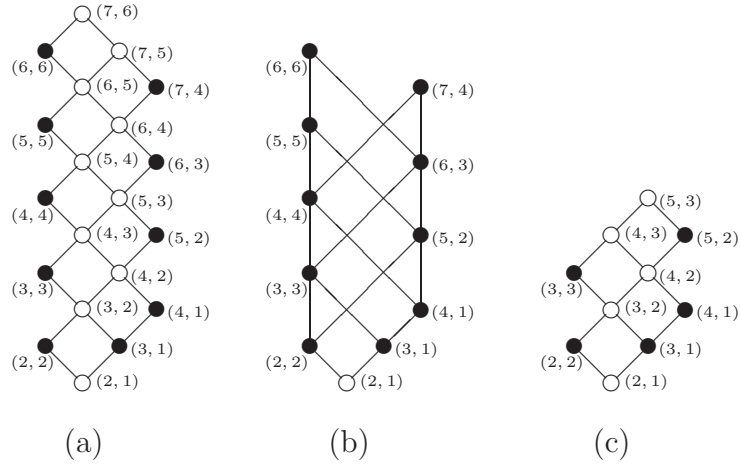
\begin{figure}[htbp]
\begin{center}
\begin{picture}(300,170)(0,-5)
\put(50,26){\circle{6}}\put(55,25){{\tiny $(2,1)$}}
\put(36,40){\circle*{6}}\put(23,32){{\tiny $(2,2)$}}
\put(64,40){\circle*{6}}\put(68,37){{\tiny $(3,1)$}}
\put(50,54){\circle{6}}\put(55,53){{\tiny $(3,2)$}}
\put(78,54){\circle*{6}}\put(82,51){{\tiny $(4,1)$}}
\put(36,68){\circle*{6}}\put(23,60){{\tiny $(3,3)$}}
\put(64,68){\circle{6}}\put(68,65){{\tiny $(4,2)$}}
\put(50,82){\circle{6}}\put(55,81){{\tiny $(4,3)$}}
\put(78,82){\circle*{6}}\put(82,79){{\tiny $(5,2)$}}
\put(36,96){\circle*{6}}\put(23,88){{\tiny $(4,4)$}}
\put(64,96){\circle{6}}\put(68,93){{\tiny $(5,3)$}}
\put(50,110){\circle{6}}\put(55,109){{\tiny $(5,4)$}}
\put(78,110){\circle*{6}}\put(82,107){{\tiny $(6,3)$}}
\put(36,124){\circle*{6}}\put(23,116){{\tiny $(5,5)$}}
\put(64,124){\circle{6}}\put(68,121){{\tiny $(6,4)$}}
\put(50,138){\circle{6}}\put(55,137){{\tiny $(6,5)$}}
\put(78,138){\circle*{6}}\put(82,135){{\tiny $(7,4)$}}
\put(36,152){\circle*{6}}\put(23,144){{\tiny $(6,6)$}}
\put(64,152){\circle{6}}\put(68,149){{\tiny $(7,5)$}}
\put(50,166){\circle{6}}\put(55,165){{\tiny $(7,6)$}}
\put(38,42){\line(1,1){10}}
\put(38,70){\line(1,1){10}}
\put(38,98){\line(1,1){10}}
\put(38,126){\line(1,1){10}}
\put(38,154){\line(1,1){10}}
\put(52,28){\line(1,1){10}}
\put(52,56){\line(1,1){10}}
\put(52,84){\line(1,1){10}}
\put(52,112){\line(1,1){10}}
\put(52,140){\line(1,1){10}}
\put(66,42){\line(1,1){10}}
\put(66,70){\line(1,1){10}}
\put(66,98){\line(1,1){10}}
\put(66,126){\line(1,1){10}}
\put(48,28){\line(-1,1){10}}
\put(48,56){\line(-1,1){10}}
\put(48,84){\line(-1,1){10}}
\put(48,112){\line(-1,1){10}}
\put(48,140){\line(-1,1){10}}
\put(62,42){\line(-1,1){10}}
\put(62,70){\line(-1,1){10}}
\put(62,98){\line(-1,1){10}}
\put(62,126){\line(-1,1){10}}
\put(62,154){\line(-1,1){10}}
\put(76,56){\line(-1,1){10}}
\put(76,84){\line(-1,1){10}}
\put(76,112){\line(-1,1){10}}
\put(76,140){\line(-1,1){10}}
%
\put(136,40){\circle*{6}}\put(119,32){{\tiny $(2,2)$}}
\put(150,26){\circle{6}}\put(155,25){{\tiny $(2,1)$}}
\put(164,40){\circle*{6}}\put(168,37){{\tiny $(3,1)$}}
\put(178,54){\circle*{6}}\put(182,51){{\tiny $(4,1)$}}
\put(136,68){\circle*{6}}\put(119,60){{\tiny $(3,3)$}}
\put(178,82){\circle*{6}}\put(182,79){{\tiny $(5,2)$}}
\put(136,96){\circle*{6}}\put(119,88){{\tiny $(4,4)$}}
\put(178,110){\circle*{6}}\put(182,107){{\tiny $(6,3)$}}
\put(136,124){\circle*{6}}\put(119,116){{\tiny $(5,5)$}}
\put(178,138){\circle*{6}}\put(182,135){{\tiny $(7,4)$}}
\put(136,152){\circle*{6}}\put(119,144){{\tiny $(6,6)$}}
\put(152,28){\line(1,1){10}}
\put(148,28){\line(-1,1){10}}
\put(138,42){\line(1,1){38}}
\put(138,70){\line(1,1){38}}
\put(138,98){\line(1,1){38}}
\put(162,42){\line(-1,1){26}}
\put(176,56){\line(-1,1){38}}
\put(176,84){\line(-1,1){38}}
\put(176,112){\line(-1,1){38}}
\put(136,40){\line(0,1){110}}
\put(178,54){\line(0,1){82}}
\put(164,40){\line(1,1){14}}
%
\put(250,26){\circle{6}}\put(255,25){{\tiny $(2,1)$}}
\put(236,40){\circle*{6}}\put(223,32){{\tiny $(2,2)$}}
\put(264,40){\circle*{6}}\put(268,37){{\tiny $(3,1)$}}
\put(250,54){\circle{6}}\put(255,53){{\tiny $(3,2)$}}
\put(278,54){\circle*{6}}\put(282,51){{\tiny $(4,1)$}}
\put(236,68){\circle*{6}}\put(223,60){{\tiny $(3,3)$}}
\put(264,68){\circle{6}}\put(268,65){{\tiny $(4,2)$}}
\put(250,82){\circle{6}}\put(255,81){{\tiny $(4,3)$}}
\put(278,82){\circle*{6}}\put(282,79){{\tiny $(5,2)$}}
\put(264,96){\circle{6}}\put(268,93){{\tiny $(5,3)$}}
\put(238,42){\line(1,1){10}}
\put(238,70){\line(1,1){10}}
\put(252,28){\line(1,1){10}}
\put(252,56){\line(1,1){10}}
\put(252,84){\line(1,1){10}}
\put(266,42){\line(1,1){10}}
\put(266,70){\line(1,1){10}}
\put(248,28){\line(-1,1){10}}
\put(248,56){\line(-1,1){10}}
\put(262,42){\line(-1,1){10}}
\put(262,70){\line(-1,1){10}}
\put(276,56){\line(-1,1){10}}
\put(276,84){\line(-1,1){10}}
\put(43,0){(a)}
\put(143,0){(b)}
\put(243,0){(c)}
\end{picture}
\caption{The planar distributive lattice for Table
 \ref{tbl:hydra-data} (a), the corresponding poset (b) and the poset
 ideal for the two-way change-point model (c).}
\label{fig:appendix-example}
\end{center}
\end{figure}
The poset
$P$ is written by the disjoint union of chains
\[
 C = \{(3,1), (4,1), (5,2), (6,3), (7,4)\} = \{a_3, a_4, a_5, a_6, a_7 \}
\]
and
\[
 D = \{(2,2), (3,3), (4,4), (5,5), (6,6) \} 
 = \{b_2, b_3, b_4, b_5, b_6\}.
\]
Note here that we are shifting the indices of $\{a_i\}, \{b_j\}$, so as
 to correspond $a_i$ to $i$-th row,  and $b_j$
to $j$-th column, respectively. 
Then for $(i,j) \in L$, we see that $i_0 = i$ and $j_0 = j$, and 
the ring homomorphism $\pi$ is
 written by $\pi(z_{ij}) = x_iy_js$ for $(i,j) \in S$ and 
$\pi(z_{ij}) = x_iy_jt$ for $(i,j) \in L\setminus S$, respectively. 

For the planar distributive lattice $L$ displayed in 
Figure \ref{fig:appendix-example}(a) and 
for the poset ideal 
$S \subset L$
displayed in 
Figure \ref{fig:appendix-example}(c), 
there are $14$ incomparable $2$-element subsets in the set $\Ac$ as
follows.
\begin{itemize}
\item
$\{\alpha, \beta, \alpha \vee \beta\} \subset S$;
\[
\begin{array}{l}
 \{(2,2),(3,1)\}, \{(2,2),(4,1)\}, \{(3,2),(4,1)\}, \{(3,3),(4,1)\},
 \{(3,3),(4,2)\},\\
 \{(3,3),(5,2)\}, \{(4,3),(5,2)\}, 
\end{array}
\]
\item
$\{\alpha, \beta, \alpha \wedge \beta\} \subset L \setminus S$;
\[
\begin{array}{l}
\{(5,5),(6,4)\}, \{(5,5),(7,4)\}, \{(6,5),(7,4)\}, \{(6,6),(7,4)\},
 \{(6,6),(7,5)\},
\end{array}\]
\item
$\alpha \in S$ and $\beta \in L \setminus S$: 
\[
 \{(4,4),(5,2)\},  \{(4,4),(5,3)\}.
\]
\end{itemize}
The set of the corresponding binomials (\ref{eqn:binom-f-a-b}) 
for these $14$ pairs coincides the set of $14$ square-free degree $2$
 moves of (\ref{eqn:minimal-MB-example}).\hspace*{\fill}$\Box$
\end{Example}

\begin{Theorem}
\label{GBtheorem}
Let $\Gc$ be the set of those binomials $f_{\alpha, \beta}$ 
with $\{\alpha, \beta\} \in \Ac$.  Then $\Gc$ is the reduced Gr\"obner basis
of $I_{(L;S)}$ with respect to $<_{\rm rev}$.
\end{Theorem}

 The proof of this theorem is in Appendix. From this theorem, we
have the following result on the Markov basis for our problem.

\begin{Theorem}\label{thm:Markov-bases}
${\cal B}^*$ is an unique minimal Markov basis for $A$ of two-way
 change-point models
 for ladder determinantal tables.
\end{Theorem}

 The uniqueness of the minimal Markov basis is from the following
known result.
\begin{Lemma}[Corollary 5.2 of
 \cite{Aoki-Hara-Takemura-2012}]\label{lem:unique-minimal}
The unique minimal Markov basis exists if and only if the set of
 indispensable moves forms a Markov basis. In this case, the set of
 indispensable moves is the unique minimal Markov basis.
\end{Lemma}

\noindent
({\it Proof of Theorem \ref{thm:Markov-bases}}.)\ 
We show ${\cal B}^*$ corresponds to the reduced Gr\"obner basis of the
 corresponding toric ideal, and therefore a Markov basis, in Theorem
 \ref{GBtheorem}. Because each element of ${\cal B}^*$ 
is an indispensable move, i.e., a difference of
 $2$-element fiber, ${\cal B}^*$ is a unique minimal Markov basis from
Lemma \ref{lem:unique-minimal}.
\hspace*{\fill}$\Box$

\section{Example}\label{sec:example}
Table \ref{tbl:hydra-data} is an example of the ladder determinantal
tables from Table 4.4-13 of \cite{Bishop-Fienberg-Holland-1975}. 
 
\begin{figure}
\begin{center}
\includegraphics[width=80mm]{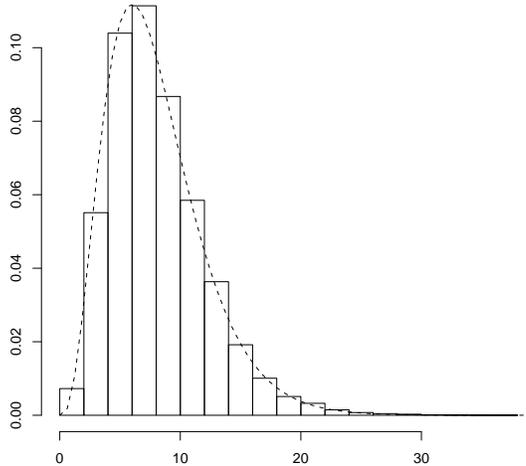}
\caption{A histogram of Pearson's goodness-of-fit $\chi^2$ generated by
 the Markov
 chain. Dotted line is the asymptotic $\chi^2_8$ distribution.}
\label{fig:hydra-histgram}
\end{center}
\end{figure}

In this experiment, annuli from donor hydra was grafted to host hydra
and observed for foot formation. The object of this experiment is to
evaluate the influence of donor and grafted annulus positions on foot
generation.
The frequencies are the cases of foot formation out
of $25$ trials, and the row and column indicate the positions
$1,\ldots,7$  from foot (position $1$) to head (position $7$) of hydra.
\begin{table}[htbp]
\begin{center}
\caption{Basal disc regeneration in hydra from 
Table 4.4-13 of
 \cite{Bishop-Fienberg-Holland-1975}}
\label{tbl:hydra-data}
\[
 \begin{array}{lc|c|c|c|c|c|c|c|}
& \multicolumn{7}{c}{\mbox{Donor annulus position}}\\
& \multicolumn{1}{c}{} & \multicolumn{1}{c}{1} & \multicolumn{1}{c}{2} &
 \multicolumn{1}{c}{3} & \multicolumn{1}{c}{4} & \multicolumn{1}{c}{5} &
 \multicolumn{1}{c}{6} & \multicolumn{1}{c}{7}\\ \cline{3-3}
& 1 & 4 & \multicolumn{6}{|c}{}\\ \cline{3-4}
& 2 & 4 & 0 & \multicolumn{5}{|c}{}\\ \cline{3-5}
\mbox{Position of graft} & 3 & 19 & 5 & 1 & \multicolumn{4}{|c}{}\\ \cline{3-6}
\mbox{in host} & 4 & 24 & 15 & 4 & 5 & \multicolumn{3}{|c}{}\\ \cline{3-7}
& \multicolumn{1}{c}{5} & \multicolumn{1}{c|}{} & 19 & 18 & 18 & 8 &
\multicolumn{2}{|c}{}\\ \cline{4-8}
& \multicolumn{1}{c}{6} & \multicolumn{2}{c|}{} & 24 & 21 & 16 & 5 &
\multicolumn{1}{|c}{}\\
  \cline{5-9}
& \multicolumn{1}{c}{7} & \multicolumn{3}{c|}{} & 23 & 22 & 8 & 1 \\
  \cline{6-9}
\end{array}
\]
\end{center}
\end{table}
For this data, though it is more natural to consider binomial sampling
model, we assume Poisson sampling model here to illustrate our method.
Then we consider the fitting of the two-way change-point model
of
\[
 B = \{(1,1),(2,1),(2,2),(3,1),(3,2),(4,1),(4,2)\}.
\]
The configuration matrix $A$ is $15\times 22$ matrix written by
\[
 A = \left(\begin{array}{c}
1000000000000000000000\\
0110000000000000000000\\
0001110000000000000000\\
0000001111000000000000\\
0000000000111100000000\\
0000000000000011110000\\
0000000000000000001111\\
1101001000000000000000\\
0010100100100000000000\\
0000010010010010000000\\
0000000001001001001000\\
0000000000000100100100\\
0000000000000000010010\\
0000000000000000000001\\
1111101100000000000000
\end{array}
\right).
\]
The fitted value of the two-way change-point model is displayed in Table
\ref{tbl:fitted-value}.
\begin{table}[htbp]
\begin{center}
\caption{Fitted value of the two-way change-point model $(i^*,j^*) =
 (4,2)$ for Table 
Table 4.4-13 of \ref{tbl:hydra-data}}
\label{tbl:fitted-value}
\[
 \begin{array}{lc|r|r|r|r|r|r|r|}
& \multicolumn{7}{c}{\mbox{Donor annulus position}}\\
& \multicolumn{1}{c}{} & \multicolumn{1}{c}{1} & \multicolumn{1}{c}{2} &
 \multicolumn{1}{c}{3} & \multicolumn{1}{c}{4} & \multicolumn{1}{c}{5} &
 \multicolumn{1}{c}{6} & \multicolumn{1}{c}{7}\\ \cline{3-3}
& 1 & 4.00 & \multicolumn{6}{|c}{}\\ \cline{3-4}
& 2 & 2.81 & 1.19 & \multicolumn{5}{|c}{}\\ \cline{3-5}
\mbox{Position of graft} & 3 & 15.94 & 6.78 & 2.28 & \multicolumn{4}{|c}{}\\ \cline{3-6}
\mbox{in host} & 4 & 28.26 & 12.03 & 4.05 & 3.67 & \multicolumn{3}{|c}{}\\ \cline{3-7}
& \multicolumn{1}{c}{5} & \multicolumn{1}{c|}{} & 19.00 & 17.17 & 15.54 & 11.29 &
\multicolumn{2}{|c}{}\\ \cline{4-8}
& \multicolumn{1}{c}{6} & \multicolumn{2}{c|}{} & 23.50 & 21.27 & 15.45 & 5.79 &
\multicolumn{1}{|c}{}\\
  \cline{5-9}
& \multicolumn{1}{c}{7} & \multicolumn{3}{c|}{} & 26.52 & 19.26 & 7.21 &
1.00 \\
  \cline{6-9}
\end{array}
\]
\end{center}
\end{table}
 As a test statistic,  we use Pearson's goodness-of-fit $\chi^2$
\[
 \chi^2 = \sum_{(i,j) \in S}\frac{(x_{ij} - m_{ij})^2}{m_{ij}},
\]
where $\Bm = (m_{ij})$ is the fitted value in Table
\ref{tbl:fitted-value}. We have $\chi^2 = 7.814$  with $8$ degrees of
freedom. 
From Theorem \ref{thm:Markov-bases}, a unique minimal Markov basis is the set
of $14$ square-free degree $2$ moves below,
\begin{equation}
\begin{array}{ccccc}
\Bz(2,3; 1,2), & \Bz(2,4; 1,2), & \Bz(3,4; 1,2), & \Bz(3,4; 1,3), & 
\Bz(3,4; 2,3),\\
\Bz(4,5; 3,4), & \Bz(4,6; 3,4), & \Bz(5,6; 3,4), & \Bz(5,6; 3,5), & 
\Bz(5,6; 4,5), \\
\Bz(5,7; 4,5), & \Bz(6,7; 4,5), & \Bz(6,7; 4,6), & \Bz(6,7; 5,6), & 
\end{array}
\label{eqn:minimal-MB-example}
\end{equation}
where $\Bz(i_1,i_2; j_1,j_2)$ is given by (\ref{eqn:basic-move-elements}). 
Using the above Markov basis, we calculate the conditional $p$ value by
the Markov chain Monte Carlo method. Starting from the observed data,
after discarding $50000$ burn-in samples, we generate $100000$ samples
from the Markov chain and have the estimate $\hat{p} = 0.46$.
Note that
the asymptotic $p$ value based on the asymptotic $\chi_8^2$ distribution of the
test statistics is $0.452$, 
which means good fitting of the asymptotic distribution for Table
\ref{tbl:hydra-data}. 
Figure \ref{fig:hydra-histgram} is a histogram of Pearson's
goodness-of-fit $\chi^2$ generated by the Markov chain, which also shows
the good fitting of the asymptotic distribution. 
 Similarly, we check the goodness-of-fits of all the two-way
change-point models for each $(i^*, j^*)$, and find that the model with 
$(i^*, j^*) = (4,2)$ is the best two-way change-point model for Table
\ref{tbl:hydra-data}, i.e., the model with the maximal estimated $p$ value.

\section{Discussion}
In this paper, we give a unique minimal Markov basis
for two-way change-point models of ladder determinantal tables. 
Our setting is an extension of two papers,
\cite{Aoki-Takemura-2005} and \cite{Hara-Takemura-Yoshida-2010}. The
two-way change-point model is an example of subtable sum problems
considered in \cite{Hara-Takemura-Yoshida-2010}, and the ladder
determinantal table is an example of incomplete contingency tables
considered in \cite{Aoki-Takemura-2005}. We  consider both
constraints at once in this paper. 
 
Our main  result is based on the theory of Gr\"obner
bases for the distributive lattice. As we see in  Section 3, the 
ladder determinantal table is treated as the distributive lattice. One
important point is that we can consider any poset ideal as the two-way
change-point models, even if it is not a rectangular shape as
(\ref{eqn:B-two-way-change-point}). Therefore
our method is also used for any $B$ as long as it corresponds to a poset
ideal of the distributive lattice. 
 
In the analysis of
two-way contingency tables, 
several extensions of the independence model
 are considered from the viewpoint of algebraic statistics. 
For example, a weakened independence model by
\cite{Carlini-Rapallo} is constructed from the set of $2\times 2$
adjacent minors.

\appendix

\section{Proof of Theorem \ref{GBtheorem}}
\begin{proof}
Once we know that $\Gc$ is a Gr\"obner basis
of $I_{(L;S)}$ with respect to $<_{\rm rev}$, it follows immediately
that $\Gc$ is reduced.
The initial monomial ${\rm in}_{<_{\rm rev}}(f_{\alpha, \beta})$
of $f_{\alpha, \beta}$ is 
${\rm in}_{<_{\rm rev}}(f_{\alpha, \beta}) = z_{\alpha}z_{\beta}$.
Let ${\rm in}_{<_{\rm rev}}(\Gc)$ denote the ideal of $K[L]$
generated by those monomials ${\rm in}_{<_{\rm rev}}(f_{\alpha, \beta})$
with $f_{\alpha, \beta} \in \Gc$.  Clearly
${\rm in}_{<_{\rm rev}}(\Gc) \subset {\rm in}_{<_{\rm rev}}(I_{(L;S)})$,
where ${\rm in}_{<_{\rm rev}}(I_{(L;S)})$ is the initial ideal of $I_{(L;S)}$
with respect to $<_{\rm rev}$.
In order to show that $\Gc$ is a Gr\"obner basis
of $I_{(L;S)}$ with respect to $<_{\rm rev}$,
by virtue of the technique \cite[Lemma 1.1]{AHH}, what we must prove is that,
for monomials $u$ and $v$, where $u \neq v$, belonging to $K[L]$ with 
$u \not\in {\rm in}_{<_{\rm rev}}(\Gc)$ and $v \not\in {\rm in}_{<_{\rm rev}}(\Gc)$,
one has $\pi(u) \neq \pi(v)$.  One can assume that $u$ and $v$ are relatively prime
and, furthermore, 
\begin{eqnarray*}
u = z_{\alpha_{1}} \cdots z_{\alpha_{p}} z_{\beta_{1}} \cdots z_{\beta_{q}},
\, \, \, \, \, 
v = z_{\alpha'_{1}} \cdots z_{\alpha'_{p}} z_{\beta'_{1}} \cdots z_{\beta'_{q}},
\end{eqnarray*}
where each $\alpha_{i} \in S$, each $\alpha'_{i} \in S$,
each $\beta_{j} \in L \setminus S$ and each $\beta'_{j} \in L \setminus S$.
Since $z_{\alpha}z_{\beta} \in {\rm in}_{<_{\rm rev}}(\Gc)$ if
$\alpha$ and $\beta$ are incomparable in $L$ with $\alpha \in S$ 
and $\beta \in L \setminus S$,
the condition that 

\noindent
\hspace{0.5cm}
$(\sharp)$ \,for each $i$ and for each $j$, one has
$\alpha_{i} < \beta_{j}$ and $\alpha'_{i} < \beta'_{j}$

\noindent
is satisfied.

If 
$\alpha_{i} \vee \alpha_{i'} \in S$,
then $\alpha_{i}$ and $\alpha_{i'}$ must be comparable in $L$.
Thus in particular, if $\alpha_{i} \vee \alpha_{i'} \in S$
for each $i$ and for each $i'$ with $1 \leq i < i' \leq p$, then 
$\{\alpha_{1}, \ldots, \alpha_{p}\}$ is a multichain of $L$.
On the other hand, suppose that
there exist $1 \leq i < i' \leq p$ with
$\alpha_{i} \vee \alpha_{i'} \in L \setminus S$.
Then, by $(\sharp)$,
for each $j$ and for each $j'$ with $1 \leq j < j' \leq q$,
one has $\beta_{j} \wedge \beta_{j'} \in L \setminus S$, so that
$\beta_{j}$ and $\beta_{j'}$ must be comparable in $L$. 
Hence $\{\beta_{1}, \ldots, \beta_{q}\}$ is a multichain of $L$.

Now, suppose that
\begin{enumerate}
\item[(i)]
for each $i$ and for each $i'$ with $1 \leq i < i' \leq p$,
one has $\alpha_{i} \vee \alpha_{i'} \in S$;
\item[(ii)]
there exist $1 \leq k < k' \leq p$ for which
$\alpha'_{k} \vee \alpha'_{k'} \in L \setminus S$.
\end{enumerate}
Then each of $\{\alpha_{1}, \ldots, \alpha_{p}\}$ and
$\{\beta'_{1}, \ldots, \beta'_{q}\}$ is a multichain of $L$.
Ignoring the variables $s$ and $t$, the toric ring $\Rc_{K}[L]$ 
introduced in \cite{Hibiring} arises.
Working in the frame of \cite{Hibiring}, 
if $u^{*} = z_{\gamma_{1}} \cdots z_{\gamma_{p+q}}$ is the standard monomial
expression of $u$ and $v^{*} = z_{\gamma'_{1}} \cdots z_{\gamma'_{p+q}}$ is 
that of $v$,
then again by $(\sharp)$ one has $|\{ i \, : \, \gamma_{i} \in S\}| \geq p$ and
$|\{ j \, : \, \gamma'_{j} \in S\}| < p$.  Hence $u^{*} \neq v^{*}$.  
Thus $\pi(u) \neq \pi(v)$.

The same argument as above shows that if we suppose
\begin{enumerate}
\item[(i')]
for each $j$ and for each $j'$ with $1 \leq j < j' \leq q$,
one has $\beta_{j} \wedge \beta_{j'} \in L \setminus S$;
\item[(ii')]
there exist $1 \leq \ell < \ell' \leq q$ for which
$\beta'_{\ell} \wedge \beta'_{\ell'} \in S$,
\end{enumerate}
then $\pi(u) \neq \pi(v)$.

Let $\pi(u) = \pi(v)$.  Then one can assume one of the following conditions:
\begin{enumerate}
\item[$(\clubsuit)$]
for each $i$ and for each $i'$ with $1 \leq i < i' \leq p$,
one has $\alpha_{i} \vee \alpha_{i'} \in S$ and 
$\alpha'_{i} \vee \alpha'_{i'} \in S$;
\item[$(\spadesuit)$]
for each $j$ and for each $j'$ with $1 \leq i < i' \leq q$,
one has $\beta_{j} \wedge \beta_{j'} \in L \setminus S$ and 
$\beta'_{j} \wedge \beta'_{j'} \in L \setminus S$.
\end{enumerate}

Suppose $(\clubsuit)$.
Then 
each of $\{\alpha_{1}, \ldots, \alpha_{p}\}$ and  
$\{\alpha'_{1}, \ldots, \alpha'_{p}\}$ is a multichain of $L$.
Hence, by $(\sharp)$ together with \cite{Hibiring},
one has $\{\alpha_{1}, \ldots, \alpha_{p}\} =  
\{\alpha'_{1}, \ldots, \alpha'_{p}\}$ as multichains of $L$.
Since $u$ and $v$ are relatively prime, one has $p = 0$.

Let $p = 0$ and $q \geq 2$. 
Let $\pi(z_{\beta_{j}}) = x_{\xi_{j}}y_{\zeta_{j}}t$ for $1 \leq j \leq q$.
Set
$\xi = \min\{ \xi_{j} \, : \, 1 \leq j \leq q \}$
and write $\zeta$ 
for the smallest integer for which there is 
$1 \leq j_{0} \leq q$ with $\pi(z_{\beta_{j_{0}}}) = x_{\xi}y_{\zeta}t$.
Then there exist $\beta'_{j_{1}}$ and $\beta'_{j_{2}}$ such that 
$\pi(\beta'_{j_{1}}) = x_{\xi}y_{j_{*}}t$ and
$\pi(\beta'_{j_{2}}) = x_{i_{*}}y_{\zeta}t$.
One has $i_{*} > \xi$ and $j_{*} > \zeta$.  
Hence $\beta'_{j_{1}} \wedge \beta'_{j_{2}} = \beta_{j_{0}}$.
Since $\beta_{j_{0}} \in L \setminus S$ and since
$\beta'_{j_{1}}$ and $\beta'_{j_{2}}$ are incomparable in $L$,
one has $z_{\beta'_{j_{1}}}z_{\beta'_{j_{2}}} \in {\rm in}_{<_{\rm rev}}(\Gc)$,
which contradicts $v \not\in {\rm in}_{<_{\rm rev}}(\Gc)$.

Finally, the same argument as above is also valid if we suppose $(\spadesuit)$.

This completes proving that 
$\Gc$ is the reduced Gr\"obner basis
of $I_{(L;S)}$ with respect to $<_{\rm rev}$.
\, \, \, \, \, \, \, \, \, \, \, \, \, \, \, \, \, \, \, \, 
\, \, \, \, \, \, \, \, \, \, \, \, \, \, \, \, \, \, \, \, 
\, \, \, \, \, \, \, \, \, \,
\end{proof}

\bibliographystyle{plain}

\begin{thebibliography}{99}
%
\bibitem{Agresti-1992}
A.~Agresti (1992).
A survey of exact inference for contingency tables. {\it Statistical
	Science}, {\bf 7}, 131--177.
%
\bibitem{Aoki-Hara-Takemura-2012}
S.~Aoki, H.~Hara and A.~Takemura (2012).
{\it Markov bases in algebraic statistics}. 
Springer Series in Statistics.
%
\bibitem{Aoki-Takemura-2005}
S.~Aoki and A.~Takemura (2005).
Markov chain Monte Carlo exact tests for incomplete two-way contingency
	tables. {\it J. Stat. Comput. Simulat.}, {\bf 75}, 787--812.
%
%
\bibitem{AHH}
A.~Aramova, J.~Herzog and T.~Hibi (2000).
Finite lattices and lexicographic Gr\"obner bases. 
{\em Europ. J. Combin.}, {\bf 21}, 431--439.
%
\bibitem{Bishop-Fienberg-Holland-1975}
Y.~M.~M.~Bishop, S.~E.~Fienberg and P.~W.~Holland (1975).
{\it Discrete multivariate analysis, Theory and applications}, The MIT
	Press, Cambridge, Massachusetts.
%
\bibitem{Carlini-Rapallo}
E.~Carlini and F.~Rapallo (2011).
A class of statistical models to weaken independence in two-way
	contingency tables. {\it Metrika}, {\bf 73}, 1--22.
%
%
%
%
%
\bibitem{Diaconis-Sturmfels-1998}
P.~Diaconis and B.~Sturmfels (1998).
Algebraic algorithms for sampling from conditional distributions. {\it
	Annals of Statistics}, {\bf 26}, 363--397.
%
\bibitem{Dobra-2003}
A.~Dobra (2003).
Markov bases for decomposable graphical models. {\it Bernoulli}, {\bf
	9}, 1093--1108.
%
%
%
\bibitem{Hastings-1970}
W.~K.~Hastings (1970).
Monte Carlo sampling methods using Markov chains and their
	applications. {\it Biometrika}, {\bf 57}, 97--109.
%
\bibitem{Hara-Takemura-Yoshida-2010}
H.~Hara, A.~Takemura and R.~Yoshida (2010).
Markov bases for subtable sum problems. {\it J. Pure Appl. Algebra},
	{\bf 213}, 1507--1521.
%
%
\bibitem{HibiRedBook}
T.~Hibi (1992).  {\em Algebraic Combinatorics on Convex Polytopes}.
Carslaw Publications, Glebe, N.S.W., Australia.
%
\bibitem{Hibiring}
T.~Hibi (1987).
Distributive lattices, affine semigroup rings and algebras with 
straightening laws. {\em Commutative Algebra and Combinatorics} 
(M. Nagata and H. Matsumura, Eds.). Advanced Studies in Pure Math., 
Volume 11, North--Holland, Amsterdam, pp.~93--109.
%
\bibitem{dojoEN}
T.~Hibi, Ed. (2013). 
{\em Gr\"{o}bner Bases: Statistics and Software Systems}. Springer.
%
\bibitem{Hirotsu-1997}
C.~Hirotsu (1997).
Two-way change-point model and its application. {\it
	Austral. J. Statist.} {\bf 39}, 205--218.
%
\bibitem{TWOWAY}
H.~Ohsugi and T.~Hibi (2009).
Two way subtable sum problems and quadratic Gr\"obner bases.
{\em Proc. Amer. Math. Soc.}, {\bf 137}, 1539--1542.
%
%
%
%
\bibitem{Pistone-Riccomagno-Wynn-2001}
G.~Pistone, E.~Riccomagno and H.~P.~Wynn (2001).
{\it Algebraic statistics: Computational commutative algebra in
	statistics}. Chapman \& Hall Ltd, Boca Raton.
%
%
%
%
\bibitem{Sturmfels-1996}
B.~Sturmfels (1996).
{\it Gr\"obner bases and convex polytopes}. In: University Lecture
	Series, {\bf 8}, American Mathematical Society, Providence, RI.
\end{thebibliography}

\end{document}